\documentclass[floatfix]{revtex4}

\usepackage{graphicx}  

\newcommand{\D}{\mbox{d}}
\newcommand{\I}{\mbox{i}}

\renewcommand{\arraystretch}{1.5}
\begin{document}
\title{Dynamics of Dissipative Quantum Systems---from Path
                       Integrals to Master Equations\footnote{
		        published in: {\em Irreversible
			   Quantum Dynamics}, F. Benatti,
			 R. Floreanini (eds.), Lecture Notes in
			 Physics 622 (Springer, New York, 2003), 165. }}
%
%
%
%
\author{Joachim Ankerhold}
\email[]{ankerhold@physik.uni-freiburg.de}
%
%
\affiliation{Physikalisches Institut, Universit{\"a}t Freiburg,
Hermann-Herder-Stra{\ss}e 3, 79104 Freiburg, Germany}


\begin{abstract}
The path integral approach offers not only an exact expression for the
nonequilibrium dynamics of dissipative quantum systems, but is
also a convenient starting point for perturbative treatments. 
An alternative way to explore the influence of friction in the
quantum realm is based on
 master equations which require, however, in one or the other
aspect approximations. Here it is discussed under which conditions and
limitations Markovian master equations can be derived from exact
path integrals thus providing a firm basis for their applicability.

\end{abstract}

\maketitle

\section{Introduction}
Quantum systems coupled to a heat bath environment can be found
almost everywhere in physics and chemistry. Recently, challenging
problems e.g.\ in realizing quantum computers, in manipulating
Bose-Einstein condensates, or in understanding low temperature
dynamics in molecular systems have triggered a substantial amount of research. 

The description of quantum dissipative processes reaches back to the
late 1960s \cite{haake1}, when it was mainly concerned with weakly damped
quantum optical systems. The conventional framework relied on
quantum Langevin or  
quantum master equations\index{master equations}. A breakthrough
beyond the limitations of the 
weak coupling approach was made in the early 1980s
\cite{weiss}. Path integral 
techniques\index{Path integral 
techniques} were shown to be powerful means to formulate the reduced
dynamics of dissipative quantum systems\index{dissipative quantum
systems} for all damping strengths, 
temperatures and bath memory times. While applications, particular in
condensed phase systems, have proven the advantage of this approach, e.g.\
the non-exponential decay of low temperature correlation functions, it
has rarely been used to put the known master equations on a firm basis
and to derive new master equations for strong damping/low
temperatures. In fact, it turned out that the quantum stochastic
process is strongly non-Markovian and intimately depends
on the initial correlations between system and bath \cite{grabert}. As
a consequence, 
``simple'' master equations do in general not exist and the known
results can be derived only in an approximate way \cite{karrlein}.
What has been essentially unexplored so far, namely 
the range of strong friction, has  
been analyzed recently in detail \cite{thorwart,ankerhold1}. It has been
shown that in this limit quantum noise is 
squeezed  and that the Wigner transform of the
reduced density obeys a  
quantum Fokker Planck equation \cite{ankerhold2}. In the low
temperature quantum domain the latter one can
be reduced to a 
generalization of the classical Smoluchowski equation
\cite{ankerhold1}. Since the path integral 
formulation is 
exact in the system-bath coupling and includes also correlated initial
states, it is basically the only way to derive these findings
consistently. 

 Thus, we are now in a position to give a more or
less comprehensive account about the validity and limitations of
Markovian master equations.
The following study is a brief attempt in this direction. It is not
intended, however, to include all developments and to 
consider the most general case. Instead, to keep things as transparent
as possible I restrict myself to (i) a single
one-dimensional continuous system with a well defined ground state coupled to a
single heat bath where (ii) the bath cut-off frequency is assumed to
be the largest frequency scale. Further, I look only for Markovian
master equations with (iii) time evolution generators independent of time
and initial preparation. 
It turns out that already within this
frame the analysis is rather subtle and reveals most of the
complexity we encounter in describing dissipative dynamics.

In Sections \ref{sec1} and \ref{sec2} I briefly introduce the path integral
approach for dissipative quantum systems. Based on this in Sec.\
\ref{sec2} some general conditions 
for the existence of Markovian master equations are derived which are then
specified in Sec.\ \ref{sec3} for weak and in Sec.\ \ref{sec4} for
strong damping, respectively.

\section{Path Integrals for Dissipative Quantum Systems}\label{sec1}

The description of classical Brownian motion  in
terms of generalized 
Langevin equations or, equivalently, in terms of Fokker-Planck
equations for  phase space distributions has a long tradition \cite{risken}. 
In contrast, the inclusion of dissipation within quantum mechanics in
a non-perturbative way has been
established only since the early 80s \cite{weiss}. In the standard
formulation one 
starts with a system+heat bath formulation 
\begin{equation}
H=H_\mathrm{S}+H_\mathrm{R}+H_\mathrm{I}\label{hamil}
\end{equation}
where the total Hamiltonian
contains a system, a reservoir (heat bath), and a system-bath interaction
part, respectively. The dynamics of
the total density matrix starting at $t=0$ from a general initial state
$W(0)$ evolves according to
\begin{equation}
W(t)= \exp(-i H t/\hbar)\, W(0)\, \exp(i H t/\hbar).\label{eq1}
\end{equation}
 The crucial point is now, that dissipation is {\em
not a priori } inherent in the system, but arises 
only if one looks on the effective impact of the bath degrees of
freedom within a reduced picture $\varrho(t)={\rm
tr}_\mathrm{R}\{W(t)\}$. The Gaussian 
statistics of the heat bath is modeled by a quasi-continuum of
harmonic oscillators bilinearly coupled with the relevant system degree of
freedom. Although the interaction between each bath degree of freedom
and the system is supposed to be weak, the overall impact of the
reservoir may cause also strong friction.
Along this reasoning two steps need to be done:  an
appropriate formulation to arrive at a reduced description has to be
found and the initial state has to be defined.  

The only non-perturbative way to deal with the elimination of the bath
degrees of freedom is to apply the path integral approach. Denoting
the degrees of freedom in coordinate  
space by $\vec{x}$ for the bath and $q$ for the system, the coordinate
representation of (\ref{eq1}) follows as 
\begin{eqnarray}
\langle q_f, \vec{x}_f|W(t)|q_f',\vec{x}_f^{\, \prime}\rangle&=&\int \D q_i
\D q_i'\, \D\vec{x}_i\D\vec{x}_i^{\,\prime }\, G( q_f,
\vec{x}_f,t,q_i,\vec{x}_i)\,\nonumber\\
&&\langle q_i, \vec{x}_i|W(0)|q_i',\vec{x}_i^{\, \prime}\rangle\,  G( q_f',
\vec{x}_f^{\, \prime},t,q_i',\vec{x}_i^{\,\prime })^\ast\label{eq1b}
\end{eqnarray}
where the $\ast$ means complex conjugation. The transition
amplitudes on the rhs are expressed as path integrals, e.g., 
\begin{equation}
G( q_f,\vec{x}_f,t,q_i,\vec{x}_i)= \int {\cal D}[q]\, {\cal
D}[\vec{x}]\, {\rm e}^{i S[q,\vec{x}]/\hbar}\label{eq1c}
\end{equation}
with the total action $S=S_\mathrm{S}+S_\mathrm{R}+S_\mathrm{I}$ according
to the three parts of the Hamiltonian (\ref{eq1}). The sum goes over
all paths running in time $t$ from $q(0)=q_i, \vec{x}(0)=\vec{x}_i$ to
$q(t)=q_f, \vec{x}(t)=\vec{x}_f$. Switching to a reduced description
is achieved by taking the trace over the bath degrees of freedom
\begin{equation}
\varrho(q_f,q_f',t)=\int d\vec{x}_f\ \langle  q_f,
\vec{x}_f|W(t)|q_f',\vec{x}_f\rangle. \label{eq1d}
\end{equation}
To carry out all integrations over the bath degrees of freedom in
(\ref{eq1c}) explicitly, the initial state must be specified.

\section{Initial State and Influence Functional}\label{sec1b}

In the ordinary Feynman Vernon\index{Feynman Vernon} theory
\cite{weiss} the initial state 
is assumed to 
be a factorizing state $W(0)=\varrho_\mathrm{S}(0)\,
Z_\mathrm{R}^{-1}\, \exp(-\beta 
H_\mathrm{R})$ ($Z_\mathrm{R}$ is the partition function of the bath) so that 
each one, 
system and equilibrated bath, lives in splendid isolation at $t=0$.
 While this assumption may be justified in the weak friction/high temperature
range, it definitely fails for moderate to strong dissipation or
lower temperature. Even the
Langevin equation is not regained in the classical limit, but differs
by initial boundary terms that may persist up to long times.
A realistic initial state reflecting the experimental situation is thus a
correlated one described by \cite{grabert}
\begin{equation}
W(0)= Z_\beta^{-1}\, \sum_i\ O_\mathrm{S}^i\, {\rm e}^{-\beta H}\,
\tilde{O}_\mathrm{S}^i\label{ini} 
\end{equation}
where $Z_\beta$ is the partition function of the total system and  the
operators $O_\mathrm{S}^i, \tilde{O}_\mathrm{S}^i$ act onto 
the system degree 
of freedom only and prepare a nonequilibrium
state. In the sequel I focus on the case where the preparation
operators depend exclusively on coordinate and refer to \cite{grabert} for the
generalization. As an example think about a position measurement with a
Gaussian slit, in which case the preparation operators are Gaussian weighted
projection operators onto position. 
Representing the
statistical operator in (\ref{ini}) as path integrals in
imaginary time we have
\begin{equation}
\langle q_i, \vec{x}_i|W(0)|q_i',\vec{x}_i^{\,
\prime}\rangle=Z_\beta^{-1}\, \lambda(q_i,q_i')\ \int {\cal
D}[\bar{q}]{\cal D}[\vec{\bar{x}}] \,
{\rm e}^{-\bar{S}[\bar{q},\bar{\vec{x}}]/\hbar} \label{eq2a}
\end{equation}
with the preparation
function $\lambda(\cdot)$ being the coordinate representation of the
preparation operators in (\ref{ini}) and the total Euclidian action
$\bar{S}=\bar{S}_\mathrm{S}+\bar{S}_\mathrm{R}+\bar{S}_\mathrm{I}$.
Paths contributing to (\ref{eq2a}) run in imaginary time 
in the interval $\hbar\beta$ from $\bar{q}(0)=q_i,
\vec{\bar{x}}(0)=\vec{x}_i$ to $\bar{q}(\hbar\beta)=q_i',
\bar{\vec{x}}(\hbar\beta)=\vec{x}_i^{\, \prime}$. 

The integrations over the bath
degrees of freedom in (\ref{eq1b}) can now together with (\ref{eq2a})
be performed exactly due to the harmonic nature of the bath. After
some tedious algebra one ends up with the 
coordinate representation of the reduced density matrix\index{reduced
density matrix} 
\begin{equation}
\varrho(q_f,q_f',t)=\int \D q_i\, \D q_i' \
J(q_f,q_f',t,q_i,q_i')\ 
{\lambda}(q_i,q_i'). \label{eq2}
\end{equation}
The propagating function $J(\cdot)$ is a threefold path integral over
the system degree of freedom only
\begin{equation}
J(q_f,q_f',t,q_i,q_i')= Z^{-1} \int {\cal D}[q]\, {\cal D}[q']\, {\cal
D}[\bar{q}]\ {\rm e}^{i\Sigma[q,q',\bar{q}]/\hbar}\label{eq2b}
\end{equation}
with $Z=Z_\beta/Z_\mathrm{R}$.  
The two real time paths $q(s)$ and $q'(s)$ connect in time $t$ the 
initial points $q_i$ and $q_i'$ with the fixed end points $q_f$ and
$q_f'$, while the imaginary time path $\bar{q}(\sigma)$ runs from $q_i$ to
$q_i'$ in the interval $\hbar\beta$.  
The contribution of each path is weighted with an effective action
$\Sigma[q,q',\bar{q}]=S_\mathrm{S}[q]-S_\mathrm{S}[q']+
i\bar{S}_\mathrm{S}[\bar{q}]+i\phi[q,q',\bar{q}]$. 
It consists of 
 the actions of the bare system in real and imaginary time,
respectively, and additional interaction contributions 
(influence functional)
\begin{eqnarray}
\phi[q,q',\bar{q}] &=&
-i\int_0^{\hbar\beta}\!\!\!\D\tau\int_0^\sigma\D\sigma\,
\bar{q}(\tau)\, K(-i\tau+i\sigma)\,
\bar{q}(\sigma)+i\int_0^{\hbar\beta}\!\!\!\D\tau \frac{\mu}{2}\, \bar{q}(\tau)^2\nonumber\\
&+&\int_0^{\hbar\beta}\!\!\!\D\tau \int_0^t\!\! \D s\,
K^*(s-i\tau)\, \bar{q}(\tau)\, x(s)-M r_i\, \gamma(s)\, x(s)\nonumber\\
& -&\int_0^t\!\! \D s\, x(s) \left[M\!  \int_0^s\!\! \D u\, \gamma(s-u)\, 
\dot{r}(u)-\frac{i}{2}\int_0^t\!\! \D u\, K'(s-u)\, 
x(u)\right].\label{phi}
\end{eqnarray}
Here, for convenience, we have introduced sum and difference real time
 paths, namely, $r(s)=[q(s)+q'(s)]/2$ and $x(s)=[q(s)-q'(s)]$.
Further, the effective impact of the bath is controlled
 by the damping kernel\index{damping kernel} 
\begin{equation}
K(\theta)=\int_0^\infty \frac{\D \omega}{\pi}\, I(\omega)\,
 \frac{\cosh[\omega(\hbar\beta/2-\I \theta)]}{\sinh(\omega\hbar\beta/2)}
 \label{eqk}
\end{equation}
where $\theta=s-\I \tau$, $0\leq s\leq t$, $0\leq \tau\leq \hbar\beta$
and $I(\omega)$ is the spectral density of the heat bath. For an
interaction term between system and bath of the form $H_\mathrm{I}=q
\xi_\mathrm{R}(\theta)$ with $\xi_\mathrm{R}(\theta) = \vec{c}\,
\vec{x}$ (coupling constants $\vec{c}$) one shows that 
the damping kernel is basically the bath autocorrelation function,
i.e.\ $K(\theta)=\langle \xi_\mathrm{R}(\theta)\,
\xi_\mathrm{R}(0)\rangle_\mathrm{R}/\hbar$. 
In real time $K(s)=K'(s)+ \I K''(s)$ is related to the
macroscopic damping kernel 
\begin{equation}
\gamma(s)=\frac{2}{M}\, \int_0^\infty \frac{\D \omega}{\pi}\,
\frac{I(\omega)}{\omega}\, \cos(\omega s)\, \label{eqd}
\end{equation}
via $K''(s)=\frac{M}{2}\D \gamma(s)/\D s$ and $K'(s)\to
M\gamma(s)/\hbar\beta$ in the classical limit $\omega_c\hbar\beta\to
0$ ($M$ is the mass of the Brownian particle). The term with
$\mu=\lim_{\hbar\beta\to 0}\hbar\beta K(0)$ in the 
first line in (\ref{phi}) gives a potential renormalization in the
Euclidian action due to 
 shifts of the minima of the bath oscillators by the coupling to the
system. The
corresponding renormalization in the real-time part of 
the action has
already been incorporated in the form it is written in (\ref{phi}).

Apparently, $\,\phi[q,q',\bar{q}]\,$ contains the bath induced non-local 
  interactions between the various system paths. Particularly, 
  it turns out that the
last term (with $K'$) acting as a Gaussian weighting factor for the
quantum fluctuations $x(s)$ suppresses coherences and gives rise to
relaxation, while the second 
last term depending on $K''(s)$ leads to dephasing during the
dynamics. The influence of initial correlations between system and
bath is contained in
the term with $K^*$ coupling real and imaginary time motion. 

 While in (\ref{eq2}) the imaginary time paths describe the initial
state, the two real time paths govern the dynamics of the reduced
system.
Accordingly, the distribution of end-points of the former and
 starting points  of the latter $q_i, q_i'$ are  weighted in
(\ref{eq2}) also by the
preparation function ${\lambda}(\cdot)$. 
In the limit $t\to 0$ one has
$J(q_f,q_f',t,q_i,q_i')\to
\varrho_\beta(q_i,q_i')\, \delta(q_f-q_i)\, 
\delta(q_f'-q_i')$ so that
\begin{equation}
\varrho(q_f,q_f',0)=
\varrho_\beta(q_i,q_i')\
{\lambda}(q_i,q_i')\ \label{eq4}  
\end{equation}
with the reduced equilibrium density matrix
$\varrho_\beta(q,q')=Z_\beta^{-1} \langle
q|{\rm tr}_\mathrm{R} \exp(-\beta H)|q'\rangle$.
In fact, this formulation reproduces in the
classical limit the generalized Langevin equation.

The nonequilibrium time evolution of a dissipative quantum system is
governed by (\ref{eq2}) together with (\ref{eq4}). The
good news is that this path integral expression is exact, also
in the system-bath coupling. The bad news is that its evaluation is
even numerically feasible only for
specific systems and certain ranges in parameter space. Since the
propagating function is
highly oscillatory,  
numerical algorithms inevitably become unstable for sufficiently long times.
In this situation ``simple'' time evolution equations are of great
practical importance and
 the path integral formulation provides a very convenient basis for
deriving them. 
 The severe problem, however, is
 the non-locality, in time, of the influence
functional\index{influence
functional}; the reduced time evolution in the time interval $s\in
[t',t]\subset [0,t]$ is 
 affected by the history of the dynamics for $0<s<t'$ and
particularly by the initial correlations between system and bath at
$s=0$. Even worse, it has been shown that {\em different} initial
densities $W(0)$ of the full compound may lead to {\em identical} reduced
initial densities $\rho(0)$, but give rise to
quite different quantum stochastic processes due to {\em different
initial correlations} \cite{grabert}.
Hence, in general a time evolution
equation $\dot{\varrho}={\cal L}\, \varrho$ of the reduced density with a
 generator ${\cal L}$ independent of time and independent of the
initial preparation does not exist.

I mention here that an alternative procedure to eliminate the bath
degrees of freedom has been developed by invoking projection operator
techniques \cite{weiss,zwanzig}. This way generalized equations of
motion in form of 
integro-differential equations for the reduced
density have been derived. It turns out, however, that this approach
is convenient only in the weak damping/high temperature range. In
contrast, the path integral formulation offers an elegant starting
point for perturbative treatments, e.g.\ semiclassical approximations,
perturbation theory {\`a} la Feynman etc., for all temperatures and
spectral bath densities. Moreover, its close relation to the formulation of
classical statistical mechanics has allowed to adopt and extend
numerical techniques like e.g.\ Monte Carlo simulations to  quantum
dissipative systems.

\section{General Conditions for Markovian Markovian Master
Equations}\label{sec2} 

As the time evolution of the reduced density matrix cannot be cast into the
form $\dot{\varrho}={\cal L}\, \varrho$ exactly, 
one may wonder under which conditions at least approximate time
evolution operators ${\cal L}$, independent of time and the
initial state, can be derived from the expression
(\ref{eq2}). Before I turn to specific cases, in this section, we
first want to formulate some general conditions. Accordingly, we
analyze the influence functional
(\ref{phi}) with respect to its non-Markovian\index{non-Markovian}
nature and with respect 
to the friction induced entanglement of the bare time evolution
operators.

For this purpose let us consider the real-time part of the damping
kernel (\ref{eqk}) in the limit of $\omega_c\to \infty$. One has
\begin{equation}
K(t)=-\frac{\pi M\gamma}{(\hbar\beta)^2}\, \frac{1}{\sinh^2(\pi
t/\hbar\beta)}\, + \I M \gamma \dot{\delta}(t).\label{kohmic}
\end{equation}
Hence, while $K''(t)$ is local in time, $K'(t)$ is not. In fact, for
$\hbar\beta\to \infty$ it only decays algebraically $K'(t)\propto
\gamma/t^2$ so that strong non-local quantum fluctuations in the influence
functional (\ref{phi}) become important and lead e.g.\ to
non-exponential long time tails for zero-temperature correlation
functions \cite{weiss}. 

One  concludes that a time evolution
operator, local in time, can only exist on a coarse grained time scale
$s\gg \hbar\beta, 1/\omega_c$. In fact,  
 a coarse graining procedure is only meaningful if
 the time scale on which the relevant 
dynamics takes place, e.g.\ the relaxation time scale
$t_r$, obeys
\begin{equation}
t_r\, \gg\, \hbar\beta,\, 1/\omega_c\, .\label{cond}
\end{equation}
Compared to the typical time scale of the bare system time evolution
$1/\omega_0$ with $\omega_0$ e.g.\ its ground state frequency, $t_r$
can be very large. For $T\to 0$ the thermal time scale exceeds all other time
scales and the above condition can never be fulfilled.

The immediate consequence of (\ref{cond}) is that the influence
functional becomes local on the coarse grained time scale.
This is fine, but it is not all we need.
Namely, the bath mediated coupling between forward and backward
time evolution and also between the time evolution and the initial
state---accounted for by the $x$ 
dependence of the influence functional (\ref{phi})---is a
genuine quantum effect, so that the $x$ paths can be interpreted to describe
quantum noise\index{quantum noise}.
  For anharmonic potential fields $V(r-x/2)$ the
effective action contains also anharmonic $x$ dependent terms meaning that
quantum noise is in general non-Gaussian. In the spirit of a classical
Kramers-Moyal expansion \cite{risken} one would expect that a time
evolution generator ${\cal L}$ of the reduced dynamics
$\rho(x_f,r_f,t)$ must be represented as an infinite  power series in 
$x_f$ and $\partial/\partial r_f$ (leading in the Wigner transform to
derivatives $\partial/\partial p$, $\partial/\partial q$) with an infinite
number of diffusion coefficients---a hopeless situation. 
For further exploring this point it is
useful to introduce the dimensionless quantity
\begin{equation}
\kappa = \left|\frac{1}{M\omega_0}\,\int_0^{t_r}\D s \int_0^{t_r}\D u \
K'(s-u)\, \right|
 = 
\frac{2\, t_r\ \gamma}{\omega_0\, \hbar\beta}\, .\label{kappa}
\end{equation}
This parameter is obtained by considering
 the last  term of (\ref{phi}), which acts as a Gaussian
weight for the $x$-paths, for
times of the order of $t_r$, putting
$x(s)=const.$, exploiting (\ref{kohmic}), and performing the coarse
graining.  Accordingly, 
the typical size of 
$|x|/\sqrt{\hbar/M\omega_0}$ is approximated to be of the order
$1/\sqrt{\kappa}$ 
and quantum noise tends to be small in domains of parameter
space where $\kappa\gg 1$.

\begin{figure}[t]
\vskip-3.5cm
\includegraphics[width=1.3\textwidth]{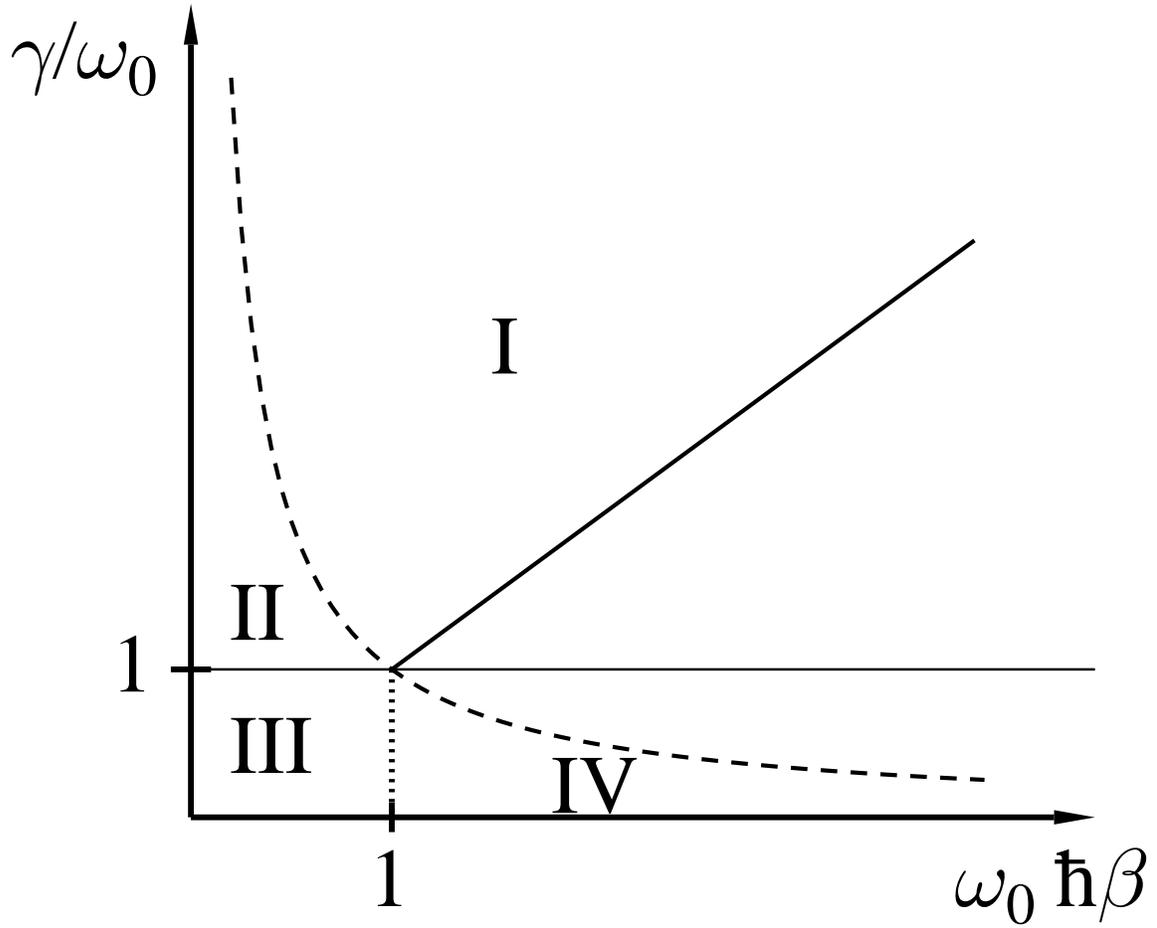}
\vskip-8.5cm
\caption[]{Sectors in parameter space  where for $\omega_c\gg
\omega_0, \gamma$ various types of master 
equations can be derived I--IV. The 
thin horizontal line separates the overdamped from the underdamped
range, the thick solid line defines the range $\gamma\hbar\beta\gg 1$
in the overdamped region, the dashed line defines
$\gamma\hbar\beta\ll 1$, and the dotted vertical line specifies
$\omega_0\hbar\beta\ll 1$ in the underdamped sector. See text for details.}
\label{fig1}
\end{figure}

In the overdamped range $\gamma/\omega_0\gg 1$ relaxation takes place
on the scale $t_r=\gamma/\omega_0^2$ so that $\kappa
=(\gamma/\omega_0)\, \gamma/\omega_0\hbar\beta$ is large under
(\ref{cond}). For sufficiently strong damping the Gaussian weighting
factor in the influence functional then causes quantum noise to be
 squeezed by friction. This allows for a semiclassical type of
approximation with generators ${\cal L}$ containing at most second
order diffusion coefficients (ranges I and II in Fig.~\ref{fig1}).
In the underdamped case $\gamma/\omega_0<1$ the dissipative system
equilibrates  on the time scale $t_r\sim 1/\gamma$ so that
(\ref{cond}) reads $\gamma\hbar\beta\ll 1$ and
$\kappa=2/\omega_0\hbar\beta$ is large as well. We this way recover
the well-known fact that
 in the
high temperature range  
$\omega_0\hbar\beta\ll 1$ quantum fluctuations are small and quantum
nonequilibrium dynamics happens to 
be close to classical nonequilibrium dynamics (range III in Fig.~\ref{fig1}).
The coarse graining condition in the underdamped regime
$\gamma\hbar\beta\ll 1$ covers also the range IV in
Fig.~\ref{fig1}. There, quantum noise is non-Gaussian and not small at
all but the
entanglement due to friction can be treated perturbatively.
To see this we estimate the contribution of kinetic terms in the bare
real-time actions to be of order $M (x/\hbar\beta)^2\,
(t_r/\hbar)=(x/\sqrt{\hbar/M\omega_0})^2 \, 1/[\gamma\omega_0
(\hbar\beta)^2]$, while the
$x^2$ term in the influence functional is
$(x/\sqrt{\hbar/M\omega_0})^2\, \kappa=(x/\sqrt{\hbar/M\omega_0})^2
2/\omega_0\hbar\beta$. Thus in the domain where the condition 
(\ref{cond}) is obeyed for $\gamma/\omega_0<1$, i.e.\ 
$\gamma\hbar\beta\ll 1$, friction terms are supposed to be much
smaller than bare kinetic contributions.

To summarize this discussion we can expect to derive
from the exact path integral expression (\ref{eq2}) approximate time
evolution operators ${\cal L}$ on a coarse grained time scale only
in the domains I--IV of Fig.~\ref{fig1}.
In the sectors I--III quantum noise is essentially
Gaussian---either due to strong friction (I) or due to high temperature
(II, III)--- while in sector IV it is non-Gaussian, but the coupling it
induces between the foward and backward time evolution of the bare
system is weak.
In the remaining parts of 
parameter space there is no reduction in a simple way possible and
one has to work either with the full path integrals or master equations
with complicated non-local integro-differential operators.
In the sequel I will focus on the ranges III, IV for weak damping
and I, II for strong friction in detail.

\section{Weak Damping Regime}\label{sec3}

It is well-known that for sufficiently high temperatures and/or
sufficiently weak friction\index{weak friction} time evolution
equations for the reduced 
dynamics can be derived \cite{haake1}. This has been done already in the late
60s motivated by the experimental progress for quantum optical
devices. Here, I discuss how these results in the {underdamped range}
can be regained from the exact path integral expression (\ref{eq2}).

In case where fluctuations tend to be small $\omega_0\hbar\beta\ll 1$
 (range III in Fig.~\ref{fig1}) master equations have been explicitly
 derived for harmonic 
 systems. 
From (\ref{eq2})  an equation of motion for $\varrho(x_f,r_f,t)$ is
 obtained which, in operator representation, coincides with the
 well-known 
Agarwal equation\index{Agarwal equation} \cite{agarwal}:\ 
 $\dot{\varrho}(t)={\cal 
 L}_{\rm Ag}\, \varrho(t)$   with
\begin{equation}
{\cal L}_\mathrm{
Ag}\, \varrho=-\frac{i}{\hbar}[H_0,\varrho]-
\frac{i\gamma}{2\hbar}[q,\{p,\varrho\}]-\frac{\gamma 
M^2\omega_0^2}{2\hbar^2}\, \langle q^2\rangle_\beta\ [q,[q,\varrho]]
\label{agarwal} 
\end{equation}
where $H_0=p^2/2M+M\omega_0^2 q^2/2$ and $[\cdot,\cdot]$\
($\{\cdot,\cdot\}$)\ denotes the commutator (anticommutator).
The equilibrium position variance is approximated to read
\begin{equation}
\langle q^2\rangle_\beta\approx\frac{\hbar}{M\omega_0}\, {\rm
coth}\left(\frac{\omega_0\hbar\beta}{2}\right).\label{q2}
\end{equation}
Since for a harmonic system quantum noise is always Gaussian, 
the above finding can be generalized to
  the wider range $\gamma\hbar\beta\ll
1$ (ranges III and IV in Fig.~\ref{fig1}).  The 
 result is an additional $p$-$q$ diffusion term, i.e.
\begin{equation}
{\cal L}_\mathrm{ ext}\, \varrho= {\cal L}_{\rm Ag}\, \varrho
+\frac{D_{pq}}{ \hbar}\,[p,[q,\varrho]]\, .\label{genagara}
\end{equation}
The  form of this additional  diffusion coefficient for the harmonic
case has been first derived from the path integral approach 
in \cite{karrlein} 
\begin{equation}
D_{pq}=\frac{1}{\hbar}\ \left(M\omega_0^2 \langle
q^2\rangle_\beta-\frac{\langle
p^2\rangle_\beta}{M}\right)\label{diffu}
\end{equation}
with 
$\langle p^2\rangle_\beta$ the equilibrium momentum variance of a
harmonic oscillator.  
For the parameter range considered here $\gamma\hbar\beta\ll 1$
($\omega_0\ll \omega_c$) one  has
\begin{equation}
\langle p^2\rangle_\beta\approx M^2 \omega_0^2 \, \langle
q^2\rangle+\frac{\hbar\gamma
M}{\pi}\left[\psi\left(1+\omega_c\hbar\beta/2\pi\right)+C\right] 
\label{p2}
\end{equation}
where $\psi(x)$ is the psi function and $C\approx 0.577$ Euler's
constant.
Accordingly, we infer from (\ref{diffu}) in ranges III/IV
\begin{equation}
D_{pq}\approx -\frac{\gamma}{\pi}\,
\left[\psi\left(1+\omega_c\hbar\beta/2\pi\right)+C \right].\label{diff}
\end{equation}
This $D_{pq}$ tends to its classical limit $D_{pq}=0$  for
$ \omega_c\hbar\beta\to 0$ with fixed
$\omega_c\gg \omega_0$. From the 
Wigner  transform of ${\cal L}_\mathrm{ ext}$ we then regain the classical
Fokker-Planck operator \cite{risken}.
For finite but still small $\omega_0\hbar\beta\ll 1$ the influence of
$D_{pq}$ is supposed to be small only, if 
additionally $\omega_c\hbar\beta\ll 1$, which gives us the precise
validity of the Agarwal equation (\ref{agarwal}) in range III.
 For lower temperatures, sectors III (partially) and IV, the $p$-$q$ diffusion
becomes  important and ${\cal L}_{\rm ext}$ with $D_{pq}$ as in (\ref{diff})
defines a master equation derived by Haake and Reibold
\cite{reibold}. Within the Nakajima-Zwanzig projection 
operator technique \cite{zwanzig} an identical equation has been
obtained within 
Redfield theory \cite{redfield}. In the limit
$\omega_c\hbar\beta\gg 1$ the expression (\ref{diff}) simplifies to
leading order to
$D_{pq}\approx (\gamma/\pi) \log(\omega_c\hbar\beta/2\pi)$.

So far I have discussed only master equations for harmonic
systems. In case of anharmonic systems and within the range
$\omega_0\hbar\beta\ll 1$ in  principle a semiclassical type of
approximation applies so that the potential field $V(r-x/2)$ can
 be expanded up to second order in $x$. This way, as mentioned above,
it is shown that 
 the $x-$paths describe 
 effectively white Gaussian noise in the reduced system. Diffusion related
 terms thus appear in ${\cal L}$ at most in second order and the
 Wigner transform of ${\cal L}$ into phase space contains at most
 second order derivatives in the phase space variables leading to a
generalized Fokker-Planck operator. 
 However, according to  the previous
section,  in the wider range III/IV a weak damping perturbation theory
of the bath induced 
entanglement is feasible anyway. Exploiting this very fact, the
nonequilibrium quantum dynamics can be cast, after an
additional coarse graining 
$s\gg 1/\omega_0$,  into
the form of the  famous Lindblad equation\index{Lindblad equation}
\cite{lindblad} 
\begin{equation}
\dot{\varrho}(t)=-\frac{i}{\hbar}[H^\prime,\varrho(t)]+\frac{1}{2\hbar}\,
\sum_l\,
[L_l\varrho(t),L_l^\dagger]+L_l,\varrho(t)L_l^\dagger]\, .\label{lindblad}
\end{equation}
Here, $H^\prime$ is a hermitian operator which must not necessarily
coincide with the bare system Hamiltonian, and the $L_l$ describe the
effective  influence of the specific heat bath. 
In case of a harmonic system and using creation and annihilation
operators the resulting master equation is also known as the
Haake-Weidlich equation \cite{haake2}. 

Before I turn to the strong friction range two remarks are in order
here. First, all time evolution operators ${\cal L}$ specified in
(\ref{agarwal}) and  (\ref{genagara}) are not of Lindblad form
(\ref{lindblad}) 
and thus do not respect complete positivity. This is not astonishing
since they  are derived based upon a coarse graining procedure.  Their
validity is restricted to a certain subspace where transient
components have  died out, while Lindblad theory itself requires a
Markovian master equation for all times. 
The path integral approach reveals explicitly that  from a physical
point of view 
this  requirement  is not necessary (see also \cite{pechukas}) and
even does not reflect the 
exact nonequilibrium dynamics.
Second, master equations have also been derived starting with
factorizing initial states (Feynman-Vernon theory), see Sec.\
\ref{sec1}, e.g.\ the 
Caldeira-Leggett master  equation\index{Caldeira-Leggett master
equation} \cite{leggett}. The problem there is 
that the  
factorizing initial state itself makes only sense in the weak damping
limit meaning that a weak friction perturbation theory within Feynman
Vernon theory is somewhat inconsistent. Particularly even the
known Fokker-Planck 
operator cannot be re-derived in the classical limit \cite{hanggi2}.
Hence, the 
Caldeira-Leggett or related master equations cannot be obtained from
(\ref{eq2}) \cite{karrlein}.

\section{Strong Friction Range}\label{sec4}

In classical physics the strong friction domain is well-known as
the Smoluchowski limit. Its characteristic property is that one has a
separation of time scales between fast equilibration of momentum and
slow relaxation in position. This way the Fokker-Planck equation for
the phase space distribution can be reduced to a Smoluchowski equation
for the marginal distribution in position space. 
For quantum dissipative systems one would think that strong friction
makes the reduced system to behave effectively more classically so that
the complicated path integral expression (\ref{eq2}) simplifies
considerably. That this is indeed the case has been analyzed in detail
only recently \cite{ankerhold1,ankerhold2} and I have given the
general argument already in 
Sec.\ \ref{sec2}. In particular,  strong dissipation\index{strong
dissipation} has quite 
a different influence on position and momentum. 

To see that in detail we define a typical
damping strength  as
\begin{equation}
\gamma\equiv \hat{\gamma}(0)=\lim_{\omega\to 0}\,
\frac{I(\omega)}{M\omega}\label{eqgam}
\end{equation}
with $\hat{\gamma}(\omega)$ the Laplace transform of the classical
damping kernel $\gamma(t)$
(\ref{eqd}). For 
instance, in the 
ohmic case $I(\omega)=M \bar{\gamma} \omega$ and also
for the more realistic Drude model
$I(\omega)=M \bar{\gamma}\omega\omega_c^2/(\omega^2+\omega_c^2)$   one finds
$\gamma=\bar{\gamma}$.
Given a typical frequency $\omega_0$ of the bare
system, e.g.\ the ground state frequency, by
strong damping we then mean 
\begin{equation}
\frac{\gamma}{\omega_0^2}\gg \hbar\beta, \frac{1}{\omega_c},
\frac{1}{\gamma}.\label{eqcon}
\end{equation}
In other words, we assume the time scale separation well-known from the
classical overdamped regime \cite{risken} and extend it to the quantum
range by incorporating the time scale for quantum fluctuations
$\hbar\beta$. 
 Correspondingly, according to the discussion in Sec.\  \ref{sec2} we
consider the dynamics (\ref{eq2}) on 
the coarse grained time scale
$s\gg \hbar\beta, \frac{1}{\omega_c},
\frac{1}{\gamma}$ and $\sigma\gg \frac{1}{\omega_c},
\frac{1}{\gamma}$.
The consequences are threefold: (a)
the strong friction 
suppresses non-diagonal elements of the
reduced density matrix during the time evolution; (b) the real-time part
$K(s)$ of the damping kernel becomes local on the coarse grained time;
(c) initial correlations as described by the first term in (\ref{phi})
survive for times of the order $\gamma/\omega_0^2$ so that factorizing
initial states cannot be used.

Following the above simplifications the path integral formulation
now allows for a perturbative treatment in the {\em strong damping
limit}. According to Sec.~\ref{sec3} the strategy is then to evaluate
the path 
integrals in the sense of a semiclassical
approximation by assuming 
self-consistently that non-diagonal elements, i.e.\
$\bar{x}=\bar{q}(\hbar\beta)-\bar{q}(0)$ and 
$x(s)$ dependent terms, remain small during the time evolution.
Hence  the effective action  
$\Sigma[r,x,\bar{q}]$ is expanded up to second order
in $\bar{x}$ of the imaginary time paths and in the
excursions $x(s)$ of the real-time path integrals. Doing so we take
sufficiently smooth potentials for granted.
 It is worthwhile to
note that we do not need to restrict the value of $\gamma\hbar\beta$
meaning that the strong damping
analysis covers a broad  temperature range from the classical
($\gamma\hbar\beta\ll 1$, range II in Fig.~\ref{fig1}) to the deep quantum 
domain ($\gamma\hbar\beta\gg 1$, range I).

Before I turn to the dynamical case, it is instructive to consider
the thermal equilibrium only. The corresponding imaginary time path
integral can  approximately be solved for an anharmonic potential and
the result is 
\begin{equation}
\varrho_\beta(\bar{x},\bar{r})= \frac{1}{Z} \, {\rm
e}^{-\beta V(\bar{r})-\langle p^2\rangle\, \bar{x}^2/2\hbar^2}\ {\rm
e}^{\Lambda\beta [ \beta V'(\bar{r})^2/2-3 V''(\bar{r})/2]}\, \label{equiden}
\end{equation}
where $\bar{r}=[\bar{q}(\hbar\beta)+\bar{q}(0)]/2$ and $Z=\int \D q 
\varrho_\beta(0,q)$. 
Further,
\begin{equation}
\Lambda=\frac{2}{M\beta}\,
\sum_{n=1}^\infty\frac{1}{\nu_n^2+\nu_n\hat{\gamma}(\nu_n)}\ \ \ \ ,\
\ \ \ \ 
\langle p^2\rangle=\frac{M}{\beta}+\frac{2
M}{\beta}
\sum_{n=1}^\infty\frac{\hat{\gamma}(\nu_n)}{\nu_n+\hat{\gamma}(\nu_n)}\,
. 
\label{coeff}
\end{equation}
Apparently, $\Lambda$ measures the typical strength of quantum
fluctuations in position 
space. In case of Drude
damping with a high frequency cut-off $\omega_c$ both $\Lambda$ and
$\langle p^2\rangle $ can be expressed in terms of $\Psi$
functions. Then, for high 
temperatures $\gamma\hbar\beta\ll 1$ we find $\Lambda\approx
\hbar^2\beta/12 M$ and $\langle p^2\rangle\approx M/\beta$.
The friction dependence appears as a genuine quantum effect for
lower temperatures and for $\gamma\hbar\beta\gg 1$ one has $\Lambda\approx 
(\hbar/M\gamma\pi) \log(\gamma\hbar\beta/2\pi)$ and $\langle p^2\rangle\approx
(M\hbar\gamma/\pi)\log(\omega_c/\gamma)$. With increasing
$\gamma$ the strong squeezing
of quantum fluctuations in
position induces enhanced quantum fluctuations in
 momentum, thus suppressing non-diagonal
elements in the density matrix. Interestingly, the probability distribution
is Gaussian in $\bar{x}$, i.e. its Wigner transform ($\bar{x}/\hbar\to
p$) Gaussian in
momentum, even at low temperatures. Anharmonic corrections in
$\bar{x}$ to the exponent are at most of order $1/\gamma^2$. Hence,
for strong friction the equilibrium density consists of a
part which in phase space takes
the form of a classical distribution, however,  with an $\hbar$ dependent
$\langle p^2\rangle$  and a part with $\Lambda$
dependent quantum corrections.  

In a similar way the dynamics can be treated.
In essence, since deviations
from diagonality 
$x(s)=q(s)-q'(s)$ remain small during the time evolution, they run 
effectively  at each instantaneous mean position $r(s)=[q(s)+q'(s)]/2$ in
a harmonic force field $V''(r) x$. Exploiting also the sluggish motion
of $r(s)$, this allows for an analytical solution which eventually
leads to a time 
evolution equation for $\varrho(q_f,q_f',t)$. After
switching to classical phase space 
$\{x_f,r_f\}\to \{p,q\}$, i.e.\ $\varrho(x_f,r_f,t)\to W(p,q,t)$, one
arrives at \cite{ankerhold2}
\begin{eqnarray}
\frac{\partial}{\partial t}\, W(p,q,t)&=& \left\{
\frac{\partial}{\partial p}\, \left[V'_{\rm eff}(q)+\gamma\,
p\right]-\frac{p}{M}\frac{\partial}{\partial q}+\gamma\, \langle p^2\rangle\,
\frac{\partial^2}{\partial p^2}\right.\nonumber\\
&&\left. +\frac{\partial^2}{\partial q\partial p}\, \left[ 1/\beta+\Lambda
\, V''(q)-\langle p^2\rangle/M\right]\right\}\ W(p,q,t)\, .\label{timeevo}
\end{eqnarray}
Here, we have introduced an effective potential $V_{\rm
eff}=V+(\Lambda/2) \, V''$. The first line on the r.h.s
coincides with a classical Fokker-Planck operator in an effective
force field, the second line describes quantum
mechanical coupled $p$-$q$ diffusion, a necessary ingredient as
already seen for the weak damping case in the previous section. In the 
high temperature limit $\gamma\hbar\beta\to 0$ the quantum
Fokker-Planck equation\index{quantum
Fokker-Planck equation} (QFP) tends to the classical Kramers equation
\cite{risken}. Small but finite $\gamma\hbar\beta\ll 1$ means in the
overdamped region also $\omega_0\hbar\beta\ll 1$ so that in case
of a {\em harmonic} potential
the QFP becomes identical to the master equation gained by Haake and
Reibold specified through (\ref{genagara}) and (\ref{diffu}).
However, while   
this known master equation is restricted (for $\gamma/\omega_0>1$) to
the range  $\gamma\hbar\beta\ll 1$, the above QFP is valid for {\em all}
$\gamma\hbar\beta$ and for anharmonic potentials as well. It is also not of 
Lindblad form due to the coarse graining procedure on which its
derivation is based.
 Of course, the equilibrium solution to (\ref{timeevo}) is given by the Wigner
transform of (\ref{equiden}). Let us 
briefly touch the question about higher order diffusion terms to
(\ref{timeevo}). For harmonic systems they do not occur so that
the QFP is in this sense exact \cite{talkner}. In case of anharmonic
potentials they result from non-Gaussian quantum
fluctuations attributed to higher than second order derivatives in
$V(q)$.  A rough 
estimate shows that
anharmonic terms in $x_f$ (leading to higher than second order
derivatives in $p$) in the crucial low temperature range
$\gamma\hbar\beta\gg 1$ are of order 
$1/[\hbar\gamma^{3/2} \log(\omega_c/\gamma)]$ compared to the leading terms. 

With the QFP at hand, we are now able to follow the reasoning of
classical mechanics and reduce it to an equation in position space--
the quantum Smoluchowski equation\index{quantum Smoluchowski equation}
(QSE).  For this purpose we employ 
the projection operator techniques 
invoked in \cite{skinner}. Along these lines one
introduces the operators ${\cal P}=f_\beta(p) \int dp$ and ${\cal
Q}=1-{\cal P}$ where $f_\beta(p)$ is the normalized momentum
distribution in equilibrium according to (\ref{equiden}).  The next steps are
 straightforward and not presented here in detail. After some
algebra one arrives to order $\Lambda/\gamma^3$ at an equation for the
position distribution 
$n(q,t)=f_\beta^{-1} \, {\cal P}W$ of the form
\begin{equation}
\frac{\partial}{\partial t} n(q,t)= \frac{1}{\gamma M}
\frac{\partial}{\partial q}\, \left\{
1+\frac{1}{M \gamma^2}\left[V''_{\rm eff}+\Lambda
V'''\right]\right\}\, L_\mathrm{ QSE}\, n(q,t)\, .\label{qse1}
\end{equation}
Here, to leading order the time evolution is determined by 
\begin{equation}
L_\mathrm{QSE}=V'_{\rm eff}+\partial/\partial q [1/\beta+\Lambda
V'']\label{qse2}
\end{equation}
 derived first in
\cite{ankerhold1}. Inertia corrections appear in curly
brackets: A classical ($\Lambda=0$)
 correction $\propto V''$ shows up,  while
quantum fluctuations enter through third and forth order derivatives
of the potential. Overdamped quantum Brownian motion in position space
thus becomes much more sensitive to the details of the potential 
profile at lower temperatures.

\section{Conclusions}

Based on the exact path integral approach I have explored to what
extent the nonequilibrium dynamics of a dissipative quantum system can
be described by a time evoution operator independent of time and
initial preparation.
The main results of this study are summarized in Table 1 together with
Fig.~\ref{fig1}. 
\begin{table}
\caption{Collection of approximate master equations
that have been derived 
within the exact path integral approach in certain domains of parameter space.}
\begin{center}
\renewcommand{\arraystretch}{1.4}
\setlength\tabcolsep{5pt}
\begin{tabular}{@{}lcccc}
\hline\noalign{\smallskip}
Master equation     & Validity$^{\mathrm{a}}$    & harmonic     &
anharmonic  &
Remarks \\
\noalign{\smallskip}
\hline
\noalign{\smallskip}
Agarwal   & III & yes & no & $\omega_c\hbar\beta\ll 1$ \\
Haake-Reibold\index{Haake-Reibold} & II, III, IV & yes & no  &derived also for 
$\omega_c>\omega_0 $ \\
Haake-Weidlich & III, IV & yes & no & coarse graining $t\gg
1/\omega_0$\\
Lindblad & III, IV & yes & yes &coarse graining $t\gg
1/\omega_0$  \\
QFP & I, II & yes &  yes & can be reduced to QSE \\
\noalign{\smallskip}
\hline
\noalign{\smallskip}
\end{tabular}
\end{center}
$^{\mathrm a}$Range of validity according to Fig.\ \ref{fig1} \\
\label{Tab}
\end{table}
The Agarwal equation, the Haake-Reibold, and the Haake-Weidlich master
equations are 
restricted to harmonic systems, where only the Haake-Reibold result is
also applicable for strong damping/high temperature. The 
Lindblad equation can be used in the weak damping domain for general
potential fields but includes an additional coarse graining on the
time scale of the bare system time evolution. For strong damping and
high as well as low temperatures only the recently found quantum
Fokker-Planck/Smoluchowski equation approximates the exact dynamics.
Combining both, Lindblad's equation and the latter one, allows to
conveniently treat dissipative quantum systems in a considerable range
of parameter space.
To obtain corresponding generators ${\cal L}$,  a coarse graining
procedure is required which renders either a semiclassical type of
approximation due to squeezed Gaussian quantum noise or a perturbative
treatment due to weak bath induced entanglement to be successful.

The area in Fig.~\ref{fig1} where no reduction to a simple time
evolution equation is possible is a real challenge for further
developments. One direction in which progress has been made uses
Hubbard-Stratonovich transformations to disentangle forward and
backward paths in the influence functional (\ref{phi}) at the expense of new
auxiliary fields. Along these lines the reduced dynamics has been expressed in
terms of  stochastic, nonlinear Schr\"odinger equation with complex
noise \cite{strunz}. In the same spirit, the quantum stochastic
process has been described to involve discontinuous jumps \cite{breuer}.
 So far, however, the applicability of related methods
is restricted 
to specific systems and/or  certain regions in parameter space. Efficient
algorithms to solve the exact path integrals (\ref{eq2}) have been
successfully developed and applied, e.g.\ real-time Monte Carlo schemes
\cite{monte} or the
so-called QUAPI-method \cite{quapi}, but the broad range of long times,
intermediate damping, low 
temperatures, and long bath memory times is still out of reach. Strong
efforts to overcome this lack are being on the way.

I am indebted to H. Grabert and P. Pechukas for many fruitful
discussions and suggestions.

\end{document}